\documentclass[prd,twocolumn,superscriptaddress]{revtex4-2}

\UseRawInputEncoding

\usepackage{graphicx, amsfonts, amsmath, amssymb, amscd,  theorem, exscale,
epic, eepic, epsfig}
\usepackage{xcolor}
\usepackage{orcidlink}

\usepackage{hyperref} %

\usepackage{makeidx}
\makeindex

\newcounter{Figure}
\theoremstyle{plain}
\theoremheaderfont{\itshape}

\theorembodyfont{\upshape}

\theoremheaderfont{\itshape} \theorembodyfont{\itshape} 
\newtheorem{results}{Main Results}
\newtheorem{state}{Statement}

\newcommand{\real}{ {\mathbb R} }

\newcommand{\ganzZahl}{ {\mathbb Z} }
\newcommand{\natZahl}{ {\mathbb N} }

\newcommand{\dlap}{\mbox{$ div \mkern -13mu / \ $}}

\newcommand{\clapa}{\mbox{$ curl \mkern -23mu / \ $}}

\newcommand{\be}{\begin{equation}}
\newcommand{\ee}{\end{equation}}
\newcommand{\bea}{\begin{eqnarray}}
\newcommand{\eea}{\end{eqnarray}}
\newcommand{\beas}{\begin{eqnarray*}}
\newcommand{\eeas}{\end{eqnarray*}}

\usepackage{xcolor}

\begin{document}

\title{Stochastic Limit of Growing Gravitational Wave Memory from Sources in the Early Universe and Astrophysical Sources}

\author{Lydia Bieri \orcidlink{0000-0002-2469-3409}}
\email{lbieri@umich.edu}
\affiliation{Department of Mathematics, University of Michigan, Ann Arbor, MI 48109-1120, USA}

%%%%%%%%%%%%%%%%%%%%%%%%%%%%%%%%%%%%%%%%%%%%%%%%%%%

\date{\today}

%%%%%%%%%%%%%%%%%%%%%%%%%%%%%%%%%%%%%%%%%%%%%%%%%%%

\begin{abstract} 
We show that the stochastic background of gravitational wave memory of growing type leads to a fractional Brownian motion increasing at the order of $t^{H}$ for large $t$ where $\frac{1}{2} < H <1$. This beats the scaling law of Brownian motion. 
In this article we investigate sources of gravitational waves in the early universe as well as in astrophysical settings. Cosmological sources may include primordial black holes or other sources immediately after the Big Bang when there were pockets of hot material, and large density fluctuations. Gravitational waves from mergers of primordial black holes produce memory. We show that due to the conditions in which these are taking place the gravitational wave memory will be increasing in time following a certain power law. Corresponding results hold for any gravitational wave memory from a cosmological source where the surrounding conditions are similar. The stochastic limit of these memories is a stochastic process growing in time faster than the $\sqrt{t}$ scaling law of Brownian motion. The latter is also typical for noise and for the limit of memory events as they have been mostly considered in the literature. In an expanding universe, the memory is enhanced by the expansion itself. Our results provide a tool to extract gravitational wave sources of this type from data using this memory signature. This would be particularly useful for the PTA data that has been already observed, answering the long-standing question on how to extract memory signals from the data. Further, the new results open up a new door to explore the conditions right after the Big Bang using the long-range dependence and further probability analysis.

\end{abstract}

\maketitle

\section{Introduction}

In the early universe right after the Big Bang, it is suggested that there were pockets of hot material with a large variety of density gradients. And the spacetime exhibited large curvature variations. 
In this article, we investigate the interactions of memory from gravitational waves from sources inside such pockets of matter in the early universe. We also address the analogous questions in astrophysical settings (thus non-cosmological sources). The common feature of the sources we are interested in, both in the cosmological as well as the astrophysical setting, is that the curvature and matter densities around the gravitational wave sources fall off ``slowly" (specified below). 
We show that gravitational waves from such sources not only give rise to a stochastic memory background, but that this stochastic memory events add up in the limit to a stochastic process that is increasing in time like $t^{\alpha}$ with $\alpha > \frac{1}{2}$. As a limiting process we obtain a fractional Brownian motion (fBM) increasing like $t^H$ with $\frac{1}{2} < H < 1$. The exact value of $H$ depends on the density of the material and geometry  around the gravitational wave sources. In particular, single memory events each increasing at the order $|t|^{1- \beta}$ for $0 < \beta < 1$ converge to fBM scaling like $t^H$ with $H = 1 - \beta/2$. In an expanding universe, this limit of memory is enhanced by the expansion itself.

A range of examples for the cosmological setting are given by primordial black holes \cite{zeldovichnvikov, hawking1971, carrhawking1974, novpolstazel1979, carrsakellariadou1999}. 
According to various studies, primordial black holes (PBH) formed during the very early phase of the universe, and estimates for mergers of such primordial black hole binaries have been performed. 
One type of scenarios of interest here consists of the interactions of memory from gravitational waves from these early PBH mergers. Such mergers occurred in regions of high density. We can think of these as being surrounded by matter that falls off with the distance in a certain fractional power law. We can think of these as ``points of concentration" embedded in these high density pockets, and these pockets having various power law density decays. 
We show that gravitational waves from mergers of such primordial black holes in the early phase not only give rise to a stochastic memory background, but that this stochastic memory events add up in the limit to a stochastic process that is increasing in time like $t^{\alpha}$ with $\alpha > \frac{1}{2}$ converging to a fBM as described above.

Other related sources consist of gravitational waves resulting from the formation of primordial black holes from high-amplitude curvature perturbations \cite{vaskverm2021}.

In the astrophysical setting, examples generating a memory background described by fBM with long-range dependence are given by binary (black hole or neutron star) mergers surrounded by clouds of neutrinos or other light matter, possibly including light dark matter.

More generally, any type of sources of gravitational waves with memory embedded in regions where the density falls off ``slowly" (specified below) give rise to a stochastic process converging to fBM with this scaling behavior. This includes cosmological sources as well as astrophysical sources.

So far, stochastic memory has been studied \cite{laskyetal2017} for sources with faster fall-off than considered here. 
Behavior deviating from the scaling of Brownian motion was computed in \cite{bmpst2024} and in \cite{uvst2025}. 
There has been an increasing amount of literature on stochastic memory. 
In the current article, we find new stochastic memory with a faster growing scaling behavior for more general sources.

Gravitational wave memory is a permanent change of the spacetime after the passage of the wave. 
The first memory was found in a linear setting \cite{zeldovichpolnarev} by Y. Zel'dovich and A. Polnarev in 1974 (ordinary memory), and in 1991 D. Christodoulou for the nonlinear Einstein equations derived a new contribution to memory (null memory) \cite{christodoulou}. The present author and D. Garfinkle showed \cite{flatmemory} that these are two different types of memory. The ordinary memory is sourced by a change of the radial-radial component of the electric part of the Weyl tensor and is due to the net change of mass and velocities in the source, whereas the null memory is sourced by the energy radiated to infinity and is computed as an integral over retarded time of the news (quadratic of the corresponding shear tensor) at null infinity of the asymptotically-flat spacetime. Typical stress-energy adds to the null memory as derived in 
\cite{1lpst1}, \cite{1lpst2}, \cite{lbdg1}, \cite{flatmemory}. 
A recent simulation was documented in \cite{ShapiroTsokarosetal2025}. 
These memories are finite and will show as a small displacement of test masses in a detector like LIGO, and in detectors using pulsar timing arrays (PTA) as a frequency 
change of pulsars' pulses. 
Memory and detectability with PTA has been studied in \cite{favata2011, seto2009, pbpusw2010, vhaalevin2010}. 
Cosmological memory was treated for de Sitter spacetime in 
\cite{bgsty1},  
for FLRW in 
\cite{twcosmo}, and 
for $\Lambda$CDM cosmology in \cite{BGYmemcosmo1}. 
It was shown that memory in de Sitter and FLRW is enhanced by a redshift factor and in $\Lambda$CDM spacetimes at cosmological distances in addition to the  enhancement by redshift, memory is altered by gravitational lensing. 
The first articles on ordinary \cite{zeldovichpolnarev} and null memory \cite{christodoulou} were followed by early works \cite{blda1, blda2, braginsky, braginskyg, will, GrPo1, thorne, thorne2, jorg}. In the meantime, the field has grown in many directions 
%\cite{1lpst1, 1lpst2, lbdg1, lbdg2, flatmemory, BGYmemcosmo1, bgsty1, twcosmo, asht1, tolwal1, winicour, Lasky1, strominger, flanagan, flanagan2, favata2011, lydia3, lydia4, lydiaalex1}. 
\cite{asht1, tolwal1, lbdg2, winicour, Lasky1, strominger, flanagan, flanagan2, lydia3, lydia4, lydiaalex1}. 
This list is not exhaustive, but rather gives an idea of various aspects of memory research. Given the vast literature on the subject, we shall concentrate on the results most relevant for the current paper. 
The present author derived \cite{lydia3, lydia4} new types of memory that are growing in retarded time for solutions of the Einstein vacuum equations as well as Einstein coupled to neutrinos. They occur in asymptotically-flat (AF) systems of slow fall-off towards infinity. These built on mathematical techniques from \cite{lydia1, lydia2}. 
We recall that future null infinity of an AF system is an ``idealization" that allows us to extract information on gravitational waves from sources described by this system. In reality, we are not observing from future null infinity, but far away from the source in the wave zone. However, the asymptotic analysis of null infinity has proven to be a most useful and precise tool to investigate gravitational waves and memory in the universe. Moreover, for standard sources, memory is a finite quantity resulting from integrating the news over infinite retarded time $u$. (The news tensor $\Xi$ is defined in appendix \ref{appe1}.) 
Even though the integral is over infinite retarded time, the actual buildup of the memory occurs during the passage of the gravitational wave burst (thus for a binary black hole merger in a fraction of a second) and looks more like a step function. Thus, in addition to the oscillatory wave signal, that goes to zero after the burst passed, there is a cumulative effect (the memory) that stays after the burst. 
Therefore, for all practical purposes, it is sufficient to compute the integral from an adequate time before the burst to an adequate time after the burst. The new types of memory derived in \cite{lydia3, lydia4} are different from this ``standard" situation in the following sense. For gravitational wave sources in AF systems with slow fall-off the memory diverges, that is for any finite retarded time $u$ the memory integral keeps growing in $u$. In this paper, we refer to this as ``growing memory". 
Thus, we distinguish between the usual radiative oscillatory signal (the actual wave burst) and the growing memory signal (non-oscillatory, cumulative) and note that the latter integral is taken over a correspondingly much larger retarded time $u$. 
In this article, we are interested in the overlap of these growing memory events at a range of finite retarded times $u$. 
Here we study memory from such sources and also in cosmological spacetimes. 
The computation of these types of memory is lined out in appendix \ref{appe1}.

The first detection of gravitational waves from binary black hole mergers in 2015 by the LIGO collaboration \cite{ligodetect1} launched a new direction of exploring the universe via gravitational waves \cite{ligodetect2, ligodetect3}. At much lower frequencies in the nHz range, using pulsar timing arrays (PTA), a stochastic gravitational wave background (GWB) was detected more recently, as reported by the North American Nanohertz Observatory for Gravitational Waves collaboration (NANOGrav) \cite{ptausananostgwb12023}, European PTA (EPTA) \cite{ptaeuropestgwb12023}, Parkes PTA (PPTA) \cite{ptaaustraliastgwb12023}, 
and Chinese PTA (CPTA) \cite{ptachinastgwb12023}, also giving supporting evidence to the quadrupolar Hellings-Downs correlation \cite{heldow1983}. 
The source for this GWB is debated in the literature. Whereas a common model suggests the main part of the signal coming from supermassive black hole binaries \cite{ptananosmbl2023}, other models refer to merging supermassive primordial black holes (PBH). A recent paper \cite{ptapbhdeptaetal20232025} finds that while homogeneously distributed PBH violate physical constraints, clustering, on the other hand, under certain conditions allows to interpret the PTA data in terms of merging PBH. 
Further suggestions, also of cosmological nature, what sources could contribute to the observed GWB are discussed in \cite{ptananoothersourcesandnew2023}. 
Yet other perspectives are given in, for instance, \cite{sasakirev2018, vaskverm2021, somaravi2025}. For more details and background, we refer to the above literature and references given therein.

In the following, after a brief review of growing memory, we derive the stochastic process and its limit process from growing memory and explore its emergence in astrophysical and cosmological settings. 
In this article, we introduce new results that are qualitatively and quantitatively very different from what has been known so far. In future work, we may explore more general data but also start fine-tuning our findings and give quantitative details.

\section{Memory in Asymptotically Flat Spacetimes} 

Most literature has studied gravitational wave sources that are stationary outside a compact set, where the gravitational wave memory is finite and of electric parity only. More general scenarios include sources that are not stationary outside a compact set, and where, as it was shown in \cite{lydia3},  \cite{lydia4}, in addition to the electric memory also magnetic memory is present and where a novel landscape of new memory structures emerges. These memories occur in spacetimes from initial data as given below in 
(\ref{afgeng})-(\ref{afgenk}) (called type (B)) and as given in 
(\ref{Oafgeng})-(\ref{Oafgenk}) (called type (G)). Another important difference to the standard settings is that the new memory structures grow in time in the (B) as well as (G) setting.  
What distinguishes data of type (B) is that the ADM energy and linear momentum are finite, and so is the null memory piece coming from the news (the purely geometric part), whereas for (G) data with $\beta < \frac{1}{2}$ the ADM energy and linear momentum are no longer bounded and null memory from news diverges as well. Here, we use the memory structures in the (B) and (G) scenarios. 

To introduce the relevant data, we first set up some notation. 
Denote by $H_0$ a spacelike hypersuface, by $\bar{g}$ a Riemannian metric on $H_0$ 
and by $k$ a symmetric $2$-tensor. Spacetimes solving the Einstein equations will be denoted by $M$ with a Lorentzian metric $g$. 
Let us consider spacetimes that are asymptotically flat of the following type.

{\itshape Initial data of type (B). (\cite{lydia1}, \cite{lydia2}).} 
We refer to an asymptotically flat initial data set as 
{\itshape (B) initial data set}, if it is 
an asymptotically flat initial data set $(H_0, \bar{g}, k)$, 
where $\bar{g}$ and $k$ are sufficiently smooth 
and 
for which there exists a coordinate system $(x^1, x^2, x^3)$ in a neighborhood of infinity such that with 
$r = (\sum_{i=1}^{3} (x^i)^2 )^{\frac{1}{2}} \to \infty$, it is:  
\bea
\bar{g}_{ij} \ & = & \ \delta_{ij} \ + \ 
o_3 \ (r^{- \frac{1}{2}}) \label{afgeng}  \\
k_{ij} \ & = & \ o_2 \ (r^{- \frac{3}{2}})    \ .    \label{afgenk}  
\eea

In \cite{lydia1}, \cite{lydia2}, under suitable smallness conditions on the initial data, global solutions to the Einstein vacuum equations were constructed, and these spacetimes themselves were shown to be asymptotically-flat and causally geodesically complete. It follows that the results along null hypersurfaces towards future null infinity are mainly independent from the smallness, and thus also hold for large data. 
We shall refer to spacetimes resulting from solving the Einstein equations with initial data as in 
(\ref{afgeng})-(\ref{afgenk}) with large data as {\itshape (B) spacetimes}. 

Exterior stability was recently shown in \cite{dshen1} by D. Shen for data where the initial metric decays to Minkowksi at infinity at a rate of $r^{- \gamma}$ for $0 < \gamma < \frac{1}{2}$, and $k$ like $r^{-1 - \gamma}$.

{\itshape Initial data of type (G). } 
We refer to an asymptotically flat initial data set as 
{\itshape (G) initial data set}, if 
$0 < \beta <1$ 
and if in the above statement equations (\ref{afgeng})-(\ref{afgenk}) are replaced by 
\bea
\bar{g}_{ij} \ & = & \ \delta_{ij} \ + \ 
o_3 \ (r^{- \beta}) \label{Oafgeng}  \\
k_{ij} \ & = & \ o_2 \ (r^{- 1- \beta})    \ .    \label{Oafgenk}  
\eea

We shall refer to spacetimes resulting from solving the Einstein equations with initial data as in (\ref{Oafgeng})-(\ref{Oafgenk}) with large data as {\itshape (G) spacetimes}. 

These fall-offs are much slower than the typically considered cases \cite{zeldovichpolnarev}, \cite{christodoulou} where the metric converges to the Minkowksi background at infinity at the order mass/r with correspondingly faster fall-off of $k_{ij}$ and the curvature components. 
New memory structures were derived for (B) and (G) spacetimes in \cite{lydia3},  \cite{lydia4}. 

The spacetime $M$ is foliated into spacelike hypersurfaces $H_t$ by a time function $t$ and into outgoing null hypersurfaces $C_u$ by an optical function $u$. (For details see \cite{lydia2, lydia4}.) Denote the $2$-surfaces of their intersections by $S_{t,u} = H_t \cap C_u$, denote by $e_4$, respectively $e_3$ the corresponding future-directed outgoing, respectively incoming null vectorfields. Let $X, Y$ be any vectors tangent to $S_{t,u}$ at a point and $\epsilon$ the area $2$-form of $S_{t,u}$. Moreover, let 
$\hat{\chi}$ denote the outgoing null shear and 
$\underline{\hat{\chi}}$ the incoming null shear. The corresponding limit of the latter at future null infinity will be denoted by $\Xi$ and is called the news tensor, see appendix \ref{appe1}. 
Then the spacetime Weyl curvature at $S_{t,u}$ decomposes into components 
$W(X, e_4, Y, e_4)$, $W(X, e_3, Y, e_3)$, 
$W(X, e_4, e_3, e_4)$, $W(X, e_3, e_3, e_4)$, 
$W(e_4, e_3, e_4, e_3)$, $W(X, Y, e_3, e_4)$. 
In (B) spacetimes the following terms were shown to have a fall-off behavior at infinity that is given by (where $r$ denotes the area radius of a local sphere and $\tau_- = \sqrt{1 + u^2}$ with $u$ corresponding to retarded time at future null infinity $\mathcal{I^+}$): 
$
W(X, e_3, Y, e_3)  =  O  ( r^{- 1} \tau_-^{- \frac{3}{2}})$, 
$W(X, e_3, e_3, e_4)  =   O  ( r^{- 2} \tau_-^{- \frac{1}{2}})$, 
$W(e_4, e_3, e_4, e_3)$,   
$W(X, Y, e_3, e_4)$, 
$ W(X, e_4, e_3, e_4)$, 
$ W(X, e_4, Y, e_4) =  o  (r^{- \frac{5}{2}}) $ , 
$\hat{\chi}  = o (r^{- \frac{3}{2}}) , 
\underline{\hat{\chi}}  =  O (r^{-1} \tau_-^{- \frac{1}{2}})
$. 
In (G) spacetimes the following terms have a fall-off behavior at infinity that is given by 
$
W(X, e_3, Y, e_3) = O ( r^{- 1}  \tau_-^{-1 - \beta})$. 
$W(X, e_3, e_3, e_4) =  O  ( r^{- 2} \tau_-^{- \beta})$. 
$W(e_4, e_3, e_4, e_3)$, 
$W(X, Y, e_3, e_4)$, 
$ W(X, e_4, e_3, e_4)$, 
$ W(X, e_4, Y, e_4) =  o  (r^{-2 - \beta})$, 
$\hat{\chi}  =  o (r^{-1 - \beta}), 
\underline{\hat{\chi}}  = O  (r^{-1} \tau_-^{- \beta})  
$. 

Gravitational wave memory in asymptotically-flat spacetimes is computed as follows (see \cite{christodoulou}, \cite{flatmemory}). 
Consider nearby geodesics being the trajectories of two test masses in free fall. Let $s$ be the initial separation in the $B$ direction. Then after the wave pulse has passed they will have a residual separation $\triangle s$ in the $A$ direction. 
From integrating the geodesic deviation equation twice and analyzing the involved Riemann curvature tensor we obtain an expression of the following form, with $m^A_{\ B}$ denoting a component of the memory tensor, 
\be \label{rsm1}
\triangle s = - \frac{s}{r} m^A_{\ B} \ 
\ee
giving the memory as the permanent displacement of nearby geodesics (respectively test masses) after the wave train passed. 
Whereas the standard memories \cite{zeldovichpolnarev}, \cite{christodoulou} build up during the time the pulse passes and leave a permanent small constant displacement of the nearby geodesics after the pulse, the new memories found in \cite{lydia3, lydia4} keep growing in retarded time after the pulse has passed. 
We recall that the leading order curvature component in the geodesic deviation equation is related to the news tensor which again is related to the shear tensor from which the memory tensor in equation (\ref{rsm1}) is computed. These quantities satisfy the Bianchi equations from which the new types of memory were derived. For a short outline of the main computation, see the appendix \ref{appe1}. 

In (B) and (G) spacetimes, the memory tensor consists of the following types of memory \cite{lydia3, lydia4}. For solutions of the Einstein-vacuum equations: 
\be \label{pqf}
P + Q + F
\ee 
for solutions of the Einstein equations coupled to corresponding stress-energy: 
\be \label{pqftetm}
P + Q + F + T_e + T_m 
\ee 
$P$ stands for the memory sourced by the $rr$ component (in some spherical coordinates) of the electric part of the Weyl tensor ($W(e_4, e_3, e_4, e_3)$), $Q$ the memory sourced by the $rr$ component of the magnetic part of the Weyl tensor, $F$ the null memory sourced by the total energy radiated away in a given direction per unit solid angle, $T_e$ the null memory sourced by the $T_{33}$ component of the stress-energy tensor (indices $_{33}$ referring to the incoming future-directed null vectorfield $e_3$), $T_m$ the null memory sourced by the corresponding curl of the stress-energy tensor. More specifics are provided in the appendix \ref{appe1}. Computations and details are given in \cite{lydia3, lydia4}. Note that in the setting of Polnarev and Zel'dovich \cite{zeldovichpolnarev} as well as Christodoulou \cite{christodoulou} there would be only $P$ (Zel'dovich-Polnarev) and $F$ (Christodoulou) in (\ref{pqf}) and the memories would be finite. Here, as derived in \cite{lydia3, lydia4}, all the memories in (\ref{pqf})-(\ref{pqftetm}) grow in time. 
The memories increase according to: $P$, $Q$, $T_e$ as $|t|^{1- \beta}$ for $0< \beta < 1$, 
$F$ as $|t|^{1- 2 \beta}$ for $0 < \beta < \frac{1}{2}$ or when replacing in (\ref{Oafgeng})-(\ref{Oafgenk}) 
the little o by big O, then $0 < \beta \leq \frac{1}{2}$, 
$T_m$ as $|t|^{1- \beta}$ for $0 < \beta \leq \frac{1}{2}$ and include $\beta = \frac{1}{2}$ only for the case when 
replacing in (\ref{Oafgeng})-(\ref{Oafgenk}) 
the little o by big O.

\section{Memory in Cosmological Spacetimes} 

In a cosmological setting, we consider the luminosity distance given by 
$d_{L} = r a (1 + z)$, where $z$ is the redshift and $a$ is the scale factor at the location of the measurement. 
On a cosmological background like de Sitter or FLRW or in a $\Lambda$CDM cosmology, $r$ is replaced by the luminosity distance $d_L$ 
\be \label{dLSm1}
\triangle s = - \frac{s}{d_L} m^A_{\ B} \ . 
\ee

Gravitational waves propagate along null geodesics of the spacetime. Therefore, we follow these waves from the source to the point of observation located in the cosmological zone, which is far from the gravitational wave source. This zone (see \cite{BGYmemcosmo1}) is characterized by the relation 
$r \gtrsim H_0^{-1}$, where $r$ is the distance from the source, and $H_0$ the Hubble parameter today. In contrast to asymptotically-flat spacetimes, where radiation can be read-off at future null infinity, there is no "null infinity" in cosmological spacetimes. Instead the notion of the cosmological zone provides a useful concept to compute radiation to be observed today in the cosmological setting.

In the following, we shall compute the results using $r$, then write them using $d_L$ instead. Then we shall make use of \cite{bgsty1}, \cite{BGYmemcosmo1} to draw conclusions for de Sitter, FLRW and $\Lambda$CDM spacetimes.

\section{Stochastic Gravitational Wave Background from Growing Memory}  

\subsection{Overview and Setting} 

In the early expanding universe, we can think of the locations where PBHs are located and where they merge as ``regions of concentration" inside pockets of matter of high density, and the density of this matter falling off at different rates. We describe these scenarios mathematically as regions around these PBHs where the solutions of the Einstein equations exhibit large concentration of curvature and energy at the center, but as we move away from it, the curvature components, metric and other geometric quantities fall off at a certain rate. Therefore, we can model this situation as many (almost) asymptotically-flat regions within a de Sitter spacetime. Consequently, results for spacetimes of types (B) and (G) describe important properties of such regions. When PBHs merge, they produce gravitational waves with memory, which will interact with other gravitational waves with memory. In this paper, we concentrate on the memory signals. Whereas waves from sources surrounded by matter that is falling-off fast enough (that is the metric falls off like ``mass/r") produce a finite memory (that is a finite small change of the spacetime) in the far field, sources within an environment where matter and the gravitational potential (the metric) fall off more slowly, namely at the level of data as in (\ref{afgeng})-(\ref{afgenk}) or data as in (\ref{Oafgeng})-(\ref{Oafgenk}), the resulting memory in the far field will be growing at the order $|t|^{1 - \beta}$ for $0 < \beta < 1$. In addition to the stochastic background of gravitational waves from such sources, there is a stochastic background of memory from the same sources. It turns out that the resulting process from memory events that are finite will converge to Brownian motion with root mean square displacement of $\sqrt{|t|}$. We observe that this is at the level of noise, and therefore will be challenging to extract from data such as observed using PTA. However, we are going to show that different results emerge from the interaction of memory events created from gravitational wave sources in regions where the matter and geometric quantities fall off more slowly in accordance with results for (B) or (G) spacetimes. As pointed out above, these memory events are growing in time $t$. We show that when  adding up the memory events where each memory is growing at the order $|t|^{1 - \beta}$ for $0 < \beta < 1$, this yields a stochastic process that will converge to fractional Brownian motion. We investigate this situation and show that the emerging fractional Brownian motion has a root mean square displacement of $t^H$ for a Hurst parameter $\frac{1}{2} < H < 1$. This is manifestly larger than the $|t|^{\frac{1}{2}}$ of standard Brownian motion with $H = \frac{1}{2}$. This compares to the long-range dependence in other stochastic processes as found in mathematical physics or finance, for instance.

Focussing on gravitational wave sources with memory according to scenarios (B) and (G), we shall first concentrate on the events growing fastest. That is, 
we are going to make use of the fact that each memory event is growing at the order 
\be \label{m1}
|t|^{1 - \beta} \ \  \mbox{for}  \ \  0 < \beta < 1 \ \ , 
\ee 
in particular for (B) this implies $\beta = \frac{1}{2}$. 

We will first consider memory that is increasing according to this law (\ref{m1}) to derive the main results \ref{The1}, and then discuss all types of memory as mentioned above in expressions (\ref{pqf})-(\ref{pqftetm}) to compute the main results \ref{The2}.

\subsection{Assumptions} 
 
First, we note that we can compute the memory after the passage of the gravitational wave train by evaluating the memory tensor at a specific point in spacetime far from the source (that is in a particular spatial location at a particular time). For each memory event from different sources we can do so. Having a large number of these events, we can describe these as a corresponding Gaussian time series taking values in $\real^n$ for $n \geq 1$. Here, we are interested only in $3+1$-dimensional spacetimes. 
We make the assumption that the components of the $n$-dimensional output at each time $t$ are independent. This will allow us to model the $n$-dimensional case as a linear combination of $1$-dimensional processes. 
One could consider also the situation where these are not independent, in which case a straightforward computation yields a formula to express the cross-covariances in the resulting processes. In future work, it will be interesting to investigate further these scenarios including non-trivial cross-covariances.

We assume that these memory events occur frequently in the early universe such that they can be described by a Gaussian time series $X = \{ X_n \}_{n \in \ganzZahl}$ with 
\be \label{var1}
var (X_1 + \cdots + X_n) = f_1(N) N^{2 - \beta} \ , \ \ N = 1, 2, \cdots , 
\ee
where $f_1(N)$ is a slowly varying function at infinity. 

We call $X = \{ X_n \}_{n \in \ganzZahl}$ a Gaussian long-range dependent (LRD) series (see \cite{taqqu1}). Note that taking the derivative of (\ref{var1}) yields the expected outcome with respect to the behavior in (\ref{m1}). See \cite{taqqu1} for a proof of convergence of Gaussian series of these types to fractional Brownian motion with Hurst parameter $\frac{1}{2} < H <1$.

\subsection{Main Results: Details} 

As we let $N \to \infty$, this series converges to fractional Brownian motion in the following sense. 

Let $X = \{ X_n \}_{n \in \ganzZahl}$ be a Gaussian long-range dependent series with variance given by (\ref{var1}). Then, as $N \to \infty$, we obtain 
\bea \label{fbh1}
& & \frac{1}{f_1(N)^{\frac{1}{2}} N^{1 - \beta/2}} \sum_{n=1}^{[Nt]} ( X_n - E[X_n]) \to B_H (t) \ \ ,  \nonumber   \\ 
& & 
 \ \ t \geq 0 \ \ ,  \label{fbh1} 
\eea
with $B_H$ being a standard fractional Brownian motion with Hurst parameter (i.e. self-similarity parameter) $H$ for $\frac{1}{2} < H <1$, where $H$ is given by 
$H = 1 - \beta/2$. 
% explain that this converges to fractional Brownian motion , see Taqqu book p. 71, prop. 2.8.7 

We obtain the following first part of the main results. (Part 2 of main results is stated below.) 

\begin{results}.  \label{The1} 
Memory events from sources whose environment is given by (B) data are described as a Gaussian LRD series $X = \{ X_n \}_{n \in \ganzZahl}$ scaling like $|t|^{\frac{1}{2}}$ and by (\ref{var1}) with variance scaling like $|t|^{\frac{3}{2}}$. This series (for large $N$ and $t$) tends to 
the limiting fractional Brownian motion (fBM) per (\ref{fbh1}) that will scale like $|t|^{\frac{3}{4}}$. (Thus, we have $\beta = \frac{1}{2}$ and $H = \frac{3}{4}$.) 
Similarly, 
memory events from sources whose environment is given by (G) data are described as a corresponding Gaussian LRD series $X = \{ X_n \}_{n \in \ganzZahl}$ 
scaling like $|t|^{1 - \beta}$ and by (\ref{var1}) with variance scaling like $|t|^{2 - \beta}$. This series (for large $N$ and $t$) converges to 
the fBM $B_H (t)$ that will scale like $|t|^H$ with $H = 1 - \beta/2$ for $0 < \beta < 1$. 
\end{results}

{\itshape Remark: }Note that for $\beta = 1$ we obtain $H = \frac{1}{2}$ which is the standard Brownian motion, that is the limiting process for the cases where the metric falls off like ``mass/r". \\ 

1) At this point, we have seen that memory events from gravitational wave sources of type (G) for a specific $\beta$ (this includes type (B) for the case when $\beta = \frac{1}{2}$) can be described by a Gaussian LRD series $X = \{ X_n \}_{n \in \ganzZahl}$, with variance given in (\ref{var1}), that converges for $N \to \infty$ to a fractional Brownian motion $B_H(t)$ with a specific $H$. Therefore, we shall from now on directly work with the emerging fractional Brownian motions $B_H(t)$ with $\frac{1}{2} < H <1$.  

2) Next, we want to allow interactions of memory signals from different sources described by scenarios (G) for various values of $\beta$. To this end, we look at the linear combination of sums of different fBMs $B_{H_k} (t)$ with $\frac{1}{2} < H_k <1$.  

3) In nature, these signals interact with signals from sources that do not produce any long-range dependent processes, but rather the limiting process is regular Brownian motion with $H = \frac{1}{2}$. Therefore, we investigate also the sum of all these types of processes described by $B_{H_k} (t)$ where now $0 < H_k <1$. Given our cosmological framework, we expect a substantial part of the involved processes to have $H_k > \frac{1}{2}$. However, for the purpose of the next statement, it is sufficient to assume that at least one of the $H_k  > \frac{1}{2}$. 

We shall make use of the following mathematical results (statement 1 and statement 2), for details of which we refer to  \cite{thaele1}, \cite{pcher1}, \cite{also1}. 
The following two statements were proven in \cite{thaele1}  by C. Th\"ale. 
Let $N \in \natZahl$ and $a_1, \cdots , a_N \in \real$ as well as $H_1, \cdots , H_N \in [0,1]$. Let 
\be \label{Y1}
Y_t := \sum_{k=1}^{N} a_k B_{H_k} (t) \ \ , \ \ t \ \in \ [0, \infty ) 
\ee
with $B_{H_k} (t)$ being independent fractional Brownian motion with Hurst parameters $H_k$ defined on some probability space. 

\begin{state}. \cite{thaele1}. 
The process $Y_t$ has the following properties. 
\begin{itemize}
\item[(a)] $Y_t$ is a Gaussian process with mean $E[Y_t] = 0$ and covariance function 
\be \label{cova1}
cov(Y_t, Y_s) = \frac{1}{2} \sum_{k=1}^N a_k^2 (t^{2H_k} + s^{2H_k} - |t-s|^{2H_k} ) 
\ee
for all $s, t, \in [0, \infty )$. 
\item[(b)] $Y_t$ has stationary increments. 
\item[(c)] Let the following be a family of scaling operators $S_{(c_1, \cdots , c_N ; H_1, \cdots , H_N)}$ with $c_1, \cdots , c_N \geq 0$, acting according to 
\beas
&& \sum_{k=1}^N f_k(t) \mapsto  S_{(c_1, \cdots , c_N ; H_1, \cdots , H_N)} \Big( \sum_{k=1}^N f_k   \Big) (t) \\ 
& & 
= 
\sum_{k=1}^N c_k^{-H_k}  f_k(c_k t) \ . 
\eeas
Then $Y_t$ is $S_{(c_1, \cdots , c_N ; H_1, \cdots , H_N)}$-invariant. That is, 
\beas
& & (S_{(c_1, \cdots , c_N ; H_1, \cdots , H_N)} Y)_t = 
\sum_{k=1}^N a_k c_k^{-H_k}  B_{H_k}(c_k t) \\ 
& & 
=
\sum_{k=1}^N a_k B_{H_k}(t) \ . 
\eeas
\item[(d)] $Y_t$ is not a Markov process except for $H_1 = \cdots = H_N = \frac{1}{2}$. 
\end{itemize}
\end{state}

\begin{state}. \cite{thaele1}. 
$Y_t$ is long-range dependent if and only if there exists some $k \in \{ 1, \cdots , N \}$ with $H_k > \frac{1}{2}$. 
\end{state}

It follows that for large values of time $t$ and $t \geq s$ the scaling behavior of the covariance $cov(Y_t, Y_s)$ is at the order of $t^{2 H_{max}}$ where $H_{max} = max_k H_k$. \\

{\itshape Details for Main Results \ref{The1}}. 
Next, we give the detailed results for the different types of memory as stated above in (\ref{pqf})-(\ref{pqftetm}).

{\it Scenario (G) with $P$ memory: }
Consider the memory generated from $P$ that is growing like $|t|^{1- \beta}$ for $0< \beta < 1$. Then we can model the occurrences of this memory as a Gaussian LRD stochastic process.  Such a process converges to fractional Brownian motion $\{ B_H (t) \}_{t \in \real}$ with $\frac{1}{2} < H <1$.  

{\it It follows} directly from assumption 1 and from the derivation above that $B_H (t)$ with $H = 1 - \beta/2$ for $0 < \beta < 1$. 
Thus, the resulting fractional Brownian motion $B_H (t)$ scales at the order of $|t|^H$. As $H > \frac{1}{2}$, these memory signals could be extracted from the stochastic background of gravitational waves or noise (that scales like regular Brownian motion at $|t|^{\frac{1}{2}}$).  

{\it Scenario (B) with $P$ memory: }
We obtain: Memory generated from $P$ that is growing like $|t|^{\frac{1}{2}}$ yields a stochastic process converging to fractional Brownian motion $\{ B_H (t) \}_{t \in \real}$ with $H = \frac{3}{4}$. 
Thus, the resulting fBM $B_H (t)$ scales at the order of $|t|^{\frac{3}{4}}$.

{\it Scenario (G) with $Q$ memory: } 
Consider the memory generated from $Q$ that is growing like $|t|^{1- \beta}$ for $0< \beta < 1$. Note that this is a rotation. Standard techniques should be applicable to extract this signal from the current data. This yields a fBM $B_H (t)$ with $H = 1 - \beta/2$ for $0 < \beta < 1$.

{\it Scenario (GT) with stress-energy and electric null memory $T_e$: } 
For spacetimes of the type (G) but coupled to corresponding stress-energy with fall-off properties in accordance with the decay of the geometric quantities (like for instance, neutrinos or very light matter), there is in addition a contribution to the electric null memory from the stress-energy component $T_{33}$ growing like $|t|^{1- \beta}$ for $0< \beta < 1$. Similar to the first situation above for $P$, we obtain the resulting fractional Brownian motion $B_H (t)$ with $H = 1 - \beta/2$ for $0 < \beta < 1$. 
Note that the corresponding limit of $T_{33}$ is integrated in time (retarded time in an asymptotically-flat manifold) and is denoted by $T_e$ in (\ref{pqftetm}).  \\

\begin{results}. \label{The2} 
\ 
\end{results} 
{\it Scenario (G) with null memory $F$: } 
Null memory $F$ from integrating the energy density (news) grows with $|t|^{1- 2 \beta}$ for $0 < \beta < \frac{1}{2}$. 
We can model the occurrences of this memory as a Gaussian LRD stochastic process with variance as in (\ref{var1}) but with $\beta$ replaced by $2 \beta$. 
Then it follows from Assumption 1 and main results \ref{The1} by replacing $\beta$ with $2 \beta$ that this stochastic process 
converges to fractional Brownian motion $\{ B_H (t) \}_{t \in \real}$ where $H = 1 - \beta$. 

{\it We observe} that this is stronger than regular Brownian motion, but less strong than the signals from $P$ and $Q$ above.

{\it Scenario (GT) with stress-energy and magnetic null memory $T_m$: } 
Consider spacetimes of the type (G), where $\beta$ is restricted to $0 < \beta \leq \frac{1}{2}$ and include $\beta = \frac{1}{2}$ only for the case when 
replacing in (\ref{Oafgeng})-(\ref{Oafgenk}) 
the little o by big O, coupled to corresponding stress-energy with fall-off properties in accordance with the decay of the geometric quantities (like for instance, neutrinos or very light matter). Then there is in addition a magnetic null memory $T_m$ sourced by the integral of the curl of the stress-energy tensor growing like $|t|^{1- \beta}$ for $0 < \beta \leq \frac{1}{2}$. 
It then follows directly that the related process converges to a fractional Brownian motion $B_H (t)$ with $H = 1 - \beta/2$ for $0 < \beta \leq \frac{1}{2}$.

{\it General, Mixed Types of Sources.} 
We find that all these memories discussed above yield fractional Brownian motions $B_H (t)$ with long-range dependence, that is 
$H > \frac{1}{2}$. Next, we combine all these fBMs $B_H (t)$ as given in (\ref{Y1}) but with $\frac{1}{2} < H_k <1$ where $1 \leq k \leq N$: 
\be \label{Y2}
Y_t = \sum_{k=1}^{N} a_k B_{H_k} (t) \ \ , \ \ t \ \in \ [0, \infty ) 
\ee
This process describes the linear combination of the independent fBMs obtained by the various types of growing memory signals. From the above  we know that $Y_t$ is a Gaussian process with mean zero. 
The covariance for $Y_t$ is given by formula (\ref{cova1}). Clearly, $Y_t$ is long-range dependent, and for large time $t$ and $t \geq s$ the behavior of $cov(Y_t, Y_s)$ is dominated by $t^{2 H_{max}}$ where $H_{max} = max_k H_k$. Therefore, for large time $t$ the process $Y_t$ scales like $t^{H_{max}}$.

If we want to allow for the regular memory effects that are not growing in time to be added, then we add regular Brownian motion (i.e. $H = \frac{1}{2}$) to the sum in (\ref{Y2}) and write for $\frac{1}{2} \leq H <1$ 
\be \label{Y3}
Y^{general}_t = \sum_{k=1}^{N} a_k B_{H_k} (t)  \ \ , \ \ t \ \in \ [0, \infty ) 
\ee
It then follows that the main scaling behavior of the process $Y^{general}_t$ does not change compared to $Y_t$. In particular, $Y^{general}_t$ is still long-range dependent, scales like $t^{H_{max}}$ and its covariance for large time $t$ and $t \geq s$ is still dominated by $t^{2 H_{max}}$. \\

\section{Fractional Brownian Motion from Memory in Asymptotically-Flat Spacetimes} 

For gravitational wave sources not at cosmological distances, the new results above hold whenever the sources are embedded in environments that can be described by spacetimes of type (G) (or (B)).

\section{Fractional Brownian Motion from Memory in Cosmological Spacetimes} 
  
In a cosmological setting, in particular for gravitational wave memory produced in mergers of PBH early on, the expansion as well as large-scale inhomogeneities may affect these formulas as follows. 
We recall that in cosmological spacetimes of the types de Sitter, FLRW and $\Lambda$CDM the memory formula is given by (\ref{dLSm1}). 
From the computations in \cite{bgsty1} adapted correspondingly to the present situation, it follows that in a de Sitter spacetime the null memories $F, T_e, T_m$ are enhanced by a factor $(1 + z)$ where $z$ is the redshift. A corresponding result for FLRW is a direct consequence from \cite{bgsty1} and \cite{twcosmo}. In \cite{BGYmemcosmo1} for $\Lambda$CDM the null memory was computed to change by a redshift factor as well as a factor due to gravitational lensing. A similar behavior is likely to hold for the new stochastic memory background. However, in article \cite{BGYmemcosmo1} a short wave-length approximation method was used. Quantitative questions on how the new results in the current paper are affected by the inhomogeneities in a $\Lambda$CDM cosmology have to be investigated in more detail which is the content of another paper.

\section{Conclusions} 

The main findings of this paper apply to the stochastic memory backgrounds for both cosmological and non-cosmological sources. 
We have found that the overlap of many memory events from gravitational wave sources not at cosmological distances (astrophysical sources), which are surrounded by neutrino clouds or other light matter, give rise to a stochastic process converging to fractional Brownian motion scaling like $t^{H}$ with $\frac{1}{2} < H < 1$. 
For sources at cosmological distances, the corresponding formulas are changed by a multiplication factor including redshift and for $\Lambda$CDM an extra factor due to gravitational lensing. Whereas in an asymptotically-flat setting (i.e. astrophysical sources) null memory is integrated with respect to retarded time at future null infinity, in a cosmological setting null memory is integrated with respect to proper time in the cosmological zone, and $r$ is replaced by $d_L$. In particular, in our model, primordial black holes in the early universe, situated in pockets of high energy or matter concentration that are falling off according to the laws of the (G) setting, give rise to a stochastic memory background scaling like $t^{H}$ with $\frac{1}{2} < H < 1$. This beats the scaling of Brownian motion. Therefore, in principle, these signals should be detectable. 
More generally, our computations show that stochastic background of any kinds of gravitational wave memory growing as discussed here will produce the limiting behavior of fractional Brownian motion with long-range dependence. 
To address further questions, it will be interesting to explore in detail the quantitative aspects for the different types of sources. 

We expect that the stochastic memory background from primordial black holes or any gravitational wave sources in the very early universe of the types  as derived in this article will carry imprints of Big Bang memory, in particular corresponding $E$-modes and $B$-modes. It will be interesting to tackle further these questions. \\

\section*{Acknowledgments} 

The author thanks the NSF. 
LB is supported by NSF grant DMS-2204182 to The University of Michigan. \\ \\

\appendix 

\section{Memory}  
\label{appe1}

For completeness and self-sufficiency, here we add a summary of the main results on growing gravitational wave memory from \cite{lydia3, lydia4} that go into formulas (\ref{rsm1})-(\ref{pqftetm}) above. See these references for details. 

In an asymptotically-flat spacetime, from the shear quantity $\underline{\hat{\chi}}$ we get the radiative 
amplitude per unit solid angle as the following limit at future null infinity 
$\Xi(u, \theta, \phi) = \lim_{C_u, r \to \infty} - \frac{1}{2} r \underline{\hat{\chi}}$, where $\theta, \phi$ denote spherical variables on $S^2$. 
The quantity $\Xi$ is also known as the news tensor. 
Then $F$ in (\ref{pqf})-(\ref{pqftetm}) above is sourced by 
$
\mathcal{F}(\cdot) = \frac{1}{2} \int_{- \infty}^{+ \infty}  \mid \Xi(u, \cdot) \mid^2  du  , 
$
where for the Einstein vacuum equations 
the energy radiated away per unit angle in a given direction is $\mathcal{F}/4\pi$. 
For the Einstein-null fluid equations describing spacetimes with neutrinos 
the energy radiated away per unit angle in a given direction is $\mathcal{F}_T/4\pi$ with 
$
\mathcal{F}_T(\cdot) = \frac{1}{2} \int_{- \infty}^{+ \infty} \left( \mid \Xi(u, \cdot) \mid^2 \ + \ 2 \pi \ \mathcal{T}_{33} (u, \cdot) \right) du   , 
$
where the second integral sources $T_e$ in (\ref{pqftetm}). 
Finally, $T_m$ in (\ref{pqftetm}) is sourced by 
the angular momentum radiated away caused by the matter 
$
\mathcal{A}_T (\cdot) = 4 \pi \int_{- \infty}^{+ \infty}  \big{(} \clapa \ \ T \big{)}^*_{34_3} (u, \cdot)  du  \ . 
$
Hereby, $\mathcal{T}_{33}$ and $\big{(} \clapa \ \ T \big{)}^*_{34_3}$ denote corresponding limits generated by the stress-energy tensor of matter. 

In \cite{lydia3, lydia4}, equations of the following form are derived for the memory tensor $m^A_{\ B}$: 
\beas
\clapa \ \  \dlap m 
& = & (\mathcal{Q}^- - \mathcal{Q}^+)   + \mathcal{A}_T      \\ 
\dlap \dlap m 
& = & (\mathcal{P}^- - \mathcal{P}^+)  - 2 \mathcal{F}_T , 
\eeas
where quantities with upper indices $+$, $-$ mean corresponding limits as retarded time $u \to \stackrel{+}{-} \infty$. 
After writing the Hodge system for these equations, the system is solved by Hodge theory on the sphere $S^2$. 
The $\mathcal{P}$ terms source $P$ and the 
$\mathcal{Q}$ terms source $Q$ in (\ref{pqf})-(\ref{pqftetm}). 
The $\mathcal{P}$ terms in the previous equations stand for the limits at future null infinity mainly including the electric Weyl curvature component $W(e_4, e_3, e_4, e_3)$ (that is the traceless part of the Riemann curvature tensor component $R(e_4, e_3, e_4, e_3)$). The $\mathcal{Q}$ terms stand for the limits at future null infinity mainly including the corresponding magnetic Weyl curvature component. These terms diverge. There is a panorama of structures stemming from these terms for which the details are given in \cite{lydia3, lydia4}. \\ \\

\end{document}